\documentstyle[prb,aps,twocolumn,epsf,floats]{revtex}

\newcommand{\de}{^{o}}

\newcommand{\beq}{\begin{equation}}
\newcommand{\eeq}{\end{equation}}
\newcommand{\bea}{\begin{eqnarray}}
\newcommand{\eea}{\end{eqnarray}}

\newcommand{\etal}{{\em et al.}}

\def\jour#1#2#3#4{{#1} {\bf #2}, #3 (#4)}
\def\tit#1#2#3#4#5{{#1} {\bf #2}, #3 (#4)}

\def\prl{Phys.\ Rev.\ Lett.\ }
\def\pr{Phys.\ Rev.\ }

\def\jmp{J.\ Math.\ Phys.\ }

\def\np{Nucl.\ Phys.\ }

\def\cd#1{c_{#1}^\dagger}
\def\fl{\Upsilon }
\def\de#1{\delta_{#1}}

\begin{document}
\draft

\twocolumn[\hsize\textwidth\columnwidth\hsize\csname @twocolumnfalse\endcsname

\title{Slow holes in the triangular Ising antiferromagnet}
 
\author{R. Moessner and S. L. Sondhi}

\address{Department of Physics, Jadwin Hall, Princeton University,
Princeton, NJ 08544, USA}
\date{\today}

\maketitle

\begin{abstract}
We consider the problem of the doped Ising antiferromagnet on the
frustrated triangular lattice in the limit where the hole kinetic
energy is much smaller than the Ising exchange. For a {\it single}
hole we prove a ``frustrated Nagaoka theorem'' showing that the ground
state is magnetized and breaks translational symmetry, in contrast to
the parent insulating state that is unmagnetized and spatially
homogenous. The extension of this physics to finite dopings depends on
the strength of a density-density coupling that is inevitably
present---we find either phase separation of the holes, or a
superconducting state that is {\it also} magnetized and breaks
translational symmetry in a feat of spatial
self-organization. Finally, we derive an effective interaction between
dilute holes at temperatures in excess of the hopping and find an
oscillatory, long-ranged form reflective of the correlations in the
underlying classical magnet which presages the breaking of
translational symmetry at zero temperature.
\end{abstract}

]

\section{Introduction}

The problem of mobile holes in a magnetic background is an especially
interesting example of charge motion in a strongly correlated electron
system. In the cases that we have in mind, the magnetism involves
local moments which reflect an underlying strong electron-electron
interaction that can produce new and interesting effects when doping
creates charge carriers. Much of the recent work on this problem has
been inspired by the cuprate superconductors, whose
physics is widely believed to be centrally connected to their genesis
as doped Mott insulators.\cite{pwa87} In this context, the central
observation is that hole motion in an ordered antiferromagnetic
background is frustrated,\cite{br} leading to the expectation that
doping\cite{fn-weak} will result in a rearrangement of the magnetic
backgound in a manner conducive to relieving this
frustration. Suggestions for what this entails include RVB (resonating
valence bond) theory,\cite{rvb} stripe formation,\cite{stripes} and,
in models with purely short-ranged interactions, phase
separation.\cite{ps} A second theme in this setting is the large spin
degeneracy of the extreme Mott insulator (e.g.\ the $U=\infty$ Hubbard
model) at half filling and the role of doping in lifting it. Perhaps
the most celebrated result along these lines is the Nagaoka
theorem,\cite{nagaoka} which established that a single hole would
completely polarize the spin background. While the situation at finite
dopings remains unsettled, the Nagaoka result does show that doping
can lift the degeneracy in striking ways.

In this paper we report some results on dilute holes introduced into a
{\it frustrated} magnetic system---the particular system studied is
the canonical example of this class, the triangular lattice Ising
antiferromagnet first studied by Wannier and Houtappel.\cite{wanhout} 
This system realises aspects
of both themes touched on above. It has local antiferromagnetic order
which leads to frustration of the hole motion. In addition however, as
this is an Ising system, the magnetic frustration leads to a finite
zero-point entropy per site, ${\cal S}$. This feature is reminiscent
of the Mott insulator cited above, but we should note that in this
case the degeneracy is generated as a cooperative effect: the Ising
model on the triangular lattice has a large number of ground states,
${\cal N}_{GS}$, because of the geometrically frustrated nature of its
magnetic interactions. In each triangle, at least one pair of spins
has to be aligned, and hence one bond frustrated. Any state with
exactly one bond per triangle is hence a ground state, and ${\cal
N}_{GS}$\ is found to scale exponentially with the number of spins,
$N$, with ${\cal S}=k_B \ln({\cal N}_{GS})/N\simeq 0.323
k_B$.\cite{wanhout}

In the following, we present a set of results on the question of if,
and how, this degeneracy is lifted upon dilute doping. To make
progress, we will assume the hole motion is slow---their kinetic
energy being taken to be much smaller than the magnetic
exchange. While Ising magnets exist, slow holes might be harder to
find. Our interest in this limit then is that a) it poses the question
of how doping interacts with a frustrated magnetic background most
cleanly, which is of theoretical interest, and b) that the ordering
patterns we find in this limit could well persist when more
enterprising holes are considered.  While more work is needed to
investigate the validity of (b) we should note concerning (a) that the
response of highly frustrated magnets to perturbations is more
generally interesting. On account of the large degeneracy, these
systems are unstable in a large number of directions, promising a wide
range of unexpected and unusual physical phenomena. Along these lines,
in a recent study along with P. Chandra, we have explored the phase
structure produced by switching on the quantum dynamics of a
transverse magnetic field instead.\cite{mcs2000}

We now turn to the Hamiltonian, $H$, we study. It is 
\bea H&=&H_t+H_J+H_\eta \nonumber \\
&=&
-t\sum_{\left< ij\right> ,\sigma} P \left( \cd{i\sigma}c_{j\sigma} + 
\cd{j\sigma}c_{i\sigma}\right) P
+J^z\sum_{\left< ij \right>} S_i^z\, S_j^z\nonumber\\
&&\ \ \ \ \ -{\eta J^z\over 4} \sum_{\left< ij
\right> }n_i n_j ,
\label{eq:ham}
\eea where $J^z>0$\ is the strength of the antiferromagnetic Ising
exchange, $t$\ is the hopping integral, and 
$c_{i\sigma}\ (\cd{i\sigma})$\
annihilates (creates) an electron at site $i$ with spin $\sigma$. 
$S_i^z= {1\over2} (\cd{i\uparrow}c_{i\uparrow} - 
\cd{i\downarrow}c_{i\downarrow})$ is the
$z$-component of the spin-1/2 operator at site $i$,  
$n_i=\sum_\sigma \cd{i\sigma}c_{i\sigma}$\
 is the total density operator and
$P$ is a projector that ensures that all sites are at most singly
occupied. Finally, the sum $\left<ij\right>$\ runs over nearest 
neighbor bonds and the sum $\sigma$\
over up and down spin orientations.  As noted earlier, we 
study the range of parameters
$t\ll J^z$, where the hopping acts as a perturbation 
to the exchange
coupling. We consider strengths of the nearest neighbour
density-density coupling of $0\leq \eta \leq 1$. In the derivation of
the rotationally invariant $t-J$\ model for $S=1/2$\ 
from the Hubbard model, such
a coupling is generated with $\eta=1$. For Ising systems, possibly
with spins greater than $1/2$, there is no such precise fixing of its
magnitude and for that reason, as well as for more general theoretical
interest, we consider a range of values for it. 

In the rest of the paper, we first prove a ``frustrated Nagaoka
theorem'', namely that the introduction of a single hole into the
ground state manifold causes the system to order into a state that is
spin polarized and breaks translational symmetry with a
$\sqrt{3}\times \sqrt{3}$-unit cell. In this state, the frustrated
bonds form a hexagonal (honeycomb) lattice on which the hole hops.
Next we consider generalizing this physics to a finite but dilute
density of holes, which turns out to depend strongly on $\eta$. For
$\eta> 1/3$, phase separation obtains.  The case $\eta<1/3$\ is
much more interesting. Here, we find a spinless Fermi liquid state
that lives on the ``hexagonal backbone'' of the single hole
problem. Further, the surrounding sites mediate an attractive
interaction that implies that the Fermi liquid is unstable to a
superconducting state in a non-zero odd angular momentum channel. The
resulting state then breaks three symmetries of the parent insulating
state at once: Ising, translation and the charge $U(1)$. Finally, we
turn to non-zero temperatures $J^z\gg T\gg t$, and show using Grassman
functional integral techniques, that there is an algebraically
decaying, entropic interaction between two holes which is oscillatory
as a function of distance and which indicates a tendency for the holes
to segregate on the hexagonal backbone.

\section{A frustrated Nagaoka theorem}

We show in this section that the ground state of our Hamiltonian
(Eq.~\ref{eq:ham}) with a single hole present is macroscopically
ordered. We will begin by establishing this in the limit where
we take $J^z=\infty$, i.e. in the degenerate perturbation theory 
problem among a set of single hole states that we identify below.
We will comment on why the conclusion is unaltered when $J^z
\gg t$ but not infinite, at the end of the proof of this first
part.

The proof will consist of first establishing a lower bound for the
energy of a one-hole state on arbitrary finite lattices and then
demonstrating that this can be saturated only on a subset of finite
lattices and that on the latter there is, up to global symmetry
operations, just one state that does so. For specificity, we will
assume that our lattices come with periodic boundary conditions.

We first identify the one-hole states that minimise the exchange energy,
noting that the term $H_\eta$\ in the Hamiltonian (Eq.~\ref{eq:ham})
is the same for any one-hole state. 
Starting from any ground state, we can only remove an electron which
experiences zero net exchange field, i.e., which has three frustrated
and three satisfied bonds. There are exponentially many such one-hole
states.

\begin{figure}
\epsfxsize=3.in
\centerline{\epsffile{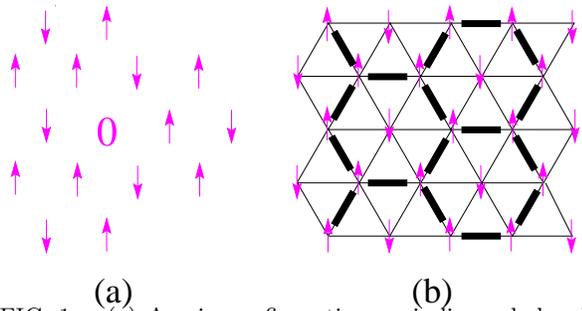}}
\caption{
(a) A spin configuration encircling a hole which allows the hole to
hop onto the three neighbouring sites occupied by up spins. 
The spins are 
denoted by arrows, the hole is represented by 0.
(b) The spin confiugration allowing the maximal coordination of three 
for a hole at any allowed site. 
Note that the sites of the up spins, on any of which a hole can reside, 
form a 
hexagonal lattice (the `hexagonal backbone', fat lines)
 on which the hole can hop.
}
\label{fig:hex}
\end{figure}

The hopping Hamiltonian connects those one-hole states which only
differ by the exchange of a hole and a neighbouring spin.
Fig.~\ref{fig:hex}a shows a spin configuration which allows the hole
to hop to three of its neighbouring sites. Up to symmetry operations,
this is the only such (local) spin configuration.  Since the hopping
integral is $t$, this means that the energy which can be gained from
$H_t$\ is at most $3t$. We note that this bound is quite general and
holds for all finite lattices with arbitrary boundary
conditions. Indeed, starting with any given one-hole state, we can
think of the all the states reached by hops as inducing a graph, whose
topology is in general far more complicated than that of the parent
lattice. The statement is that regardless of the complexity, the graph
has a maximum local coordination number of 3 and that the nearest
neighbor hopping problem on it has at best an energy of $-3t$.

This energy gain can only be realised if any state reached after one
hop again permits hoppping in three directions. Demanding that this be
the case completely constrains the parent spin configuration to the
one depicted in Fig.~\ref{fig:hex}b; the other five states that
satisfy this condition are obtained using translations or Ising
reversal.  In this configuration, the hole can occupy any site with a
frustrated bond. With periodic boundary conditions, it is clear that
not all finite lattices accomodate this pattern, but that there
is a subset of arbitrarily large size which does---these are the 
lattices with linear sizes that are integer
multiples of 3.  In the following we construct our desired
one-hole eigenstate $\left. |h \right>$\ with energy $-3t$ on this
subset.

Define $\left. |G \right> = \prod_i
\cd{i,\sigma{(i)}}\left. |0\right>$, where $\sigma{(i)}$\ is the spin
at site $i$ in the hexagonal state depicted in Fig.~\ref{fig:hex}b,
and $\left. |0\right>$\ is the empty lattice; the ordering of the
operators is immaterial in what follows. Next, define the state with a
hole on site $n$\ as $\left. |n \right>=(-1)^S c_{n,\uparrow}
\left. |G \right>$. Here, $S(n)=0\ (1)$\ if the hole is created on
sublattice A (B) of the bipartite hexagonal lattice 
defined by the frustrated
bonds. The matrix elements $\left< m|H_t|n\right>$\ vanish unless $m$\
and $n$\ refer to neighbouring sites on the hexagonal lattice, in
which case \bea \left< m|H_t|n\right> &=&\left< m\right|
-t\sum_{ij,\sigma}\cd{i\sigma}c_{j\sigma}\left| n\right> \nonumber\\
&=&-t \left< m\right| \cd{n\uparrow}c_{m\uparrow}\left| n\right>
\nonumber\\ &=&-t (-1)^{(0+1)}\left< G\right|
\cd{m\uparrow}\cd{n\uparrow} c_{m\uparrow}c_{n\uparrow}\left| G\right>
\nonumber\\ &=&-t\left< G\right| \cd{n\uparrow}c_{n\uparrow}
\cd{m\uparrow}c_{m\uparrow}\left| G\right>=-t\ . \eea

Now, let $\left. |h \right>=\sum_{i=1}^{2N/3}\left| n\right>$. For
this state, we have $H_t\left. |h \right>=-3t\left. |h \right>$\ since
$H_t$\ connects, with amplitude $-t$, each single-hole component of
the wavefunction to the three components with the hole on a
neighbouring site. Note that this is, in essence, the ground state of a
{\it spinless} hole hopping on a hexagonal lattice.

Unlike the ensemble of undiluted ground states, which has neither
long-range order nor a net magnetisation, $\left. |h \right>$\
incorporates long-range spin order.\cite{fn-tgoestozero} There is a
three sublattice ($\sqrt{3}\times\sqrt{3}$) ordering pattern with the
spin on two sublattices pointing up and down on the third.  The
magnetisation induced by adding the single hole is therefore $N/3$. It
is interesting to note that this ordering pattern would also have been
generated by an explicitly symmetry-breaking magnetic field pointing
along the Ising axis and that the $\sqrt{3}\times\sqrt{3}$\ structure
is also picked out by a {\it transverse} field.\cite{mcs2000}

Finally, we ask whether the conclusion at $J^z = \infty$\ is altered
when $J^z$\ is made finite. While we do not have a proof of this, we
believe that this does not happen. Conceptually, it is useful to
imagine generating an effective Hamiltonian in the space of all
one-hole states by including fluctuations outside this space
perturbatively in powers of $t/J^z$. Such an expansion will not
invalidate the conclusion reached inside the hexagonal backbone
states---it will only modify the state $\left. |h \right>$\
perturbatively, preserving its symmetry characteristics at small
$t/J^z$. The real danger is that a state in a different sector that
has an energy $\epsilon=-3t + \theta(L)$\ 
for a system of linear dimension $L$,
when $J^z=\infty$, with $\lim_{L\rightarrow\infty}\theta(L)=0^+$,
will acquire a perturbative gain of $O\left(
[t/J^z]^n\right)$\ for some $n$, in excess of the state $\left. |h
\right>$, thereby invalidating our conclusion at sufficiently large
$L$.

To gain insight into why this is unlikely to happen, consider the
energetics at $O\left([t/J^z]^2\right)$. At this order, the hole can
hop to a site where the net exchange field does not vanish and,
depending on the spin configuration, either be forced to return to its
starting site or hop to a different one. In terms of the induced graph
mentioned above, the former corresponds to an on-site potential energy
term, while the latter generates a new, or modifies the strength of an
existing, edge (hopping amplitude). In the hexagonal backbone states,
only the first option is available, yielding an effective potential
energy of $-{3 t^2/J^z}$\ at each site. If we consider altering the
state near a particular site to generate more favourable potential or
hopping matrix elements to $O\left([t/J^z]^2\right)$, this invariably
requires sacrificing a hopping element of $O(t)$.  We believe that
this points to a certain rigidity of the hexagonal backbone which will
operate to exclude similar subtleties at higher orders as well.
Needless to say, a proof of this assertion would be 
desirable.\cite{fn-contrast}


\section{Small density of holes and leading 
$\lowercase{t}/J^{\lowercase{z}}$ corrections}

We begin with the case $\eta=0$. First, we need to find the allowed
many hole states at low dilution (and when $t/J^z \rightarrow 0$). As
before, we need to remove those electrons experiencing zero net
exchange field since this ensures that the leading term of the
Hamiltonian, $H_J$, continues to be optimised. This implies that one
cannot remove two neighouring electrons: this would leave a triangle
with only one occupied site and hence no contribution to the magnetic
energy, whereas the optimal states are those in which the three bonds
of each triangle have total magnetic energy $-J/4$.

We showed in the last section that the kinetic energy selects a
subset of the one-hole states in which the hole is maximally
delocalized. We believe that at sufficiently low dopings, this logic
continues to operate and the low-lying eigenstates of $H$\
will be constructed out of the many-hole states obtained by removing
electrons from the hexagonal backbone of the state depicted in
Fig~\ref{fig:hex}b. (However, we are no longer able to prove this.)
This state continues to be favourable as it allows the holes to be
maximally mobile by having coordination three everywhere. The only
restriction being placed on the holes is not to sit immediately next
to each other, which is a demand present  for all possible
background states and one that should not be too onerous at low
dopings.

With these restrictions, the problem reduces to that of spinless
fermions hopping on a hexagonal lattice with amplitude $t$ and with a
nearest neighbor hard core repulsion. At low densities, this is a well
studied problem and known to be a Fermi liquid.\cite{fl} In this
fashion, we find that the ground state of the dilute hole problem is
magnetized, breaks translational symmetry {\it and} is a two
dimensional Fermi liquid.

Now this is not quite true, for the Fermi liquid suffers from the
well-known Kohn-Luttinger superconducting instability \cite{kohn-lutt}
and so strictly at $T=0$ the system will break the charge $U(1)$
symmetry as well---thereby breaking three symmetries of the parent
insulating state at once!  While this happens as $t/J^z \rightarrow
0$, we show next that for $t/J^z$ large but not infinite, the magnetic
background of the hexagonal conducting lattice mediates an attractive
interaction that also drives a superconducting instability---at low
densities this is a much larger effect than the purely ``internal''
Kohn-Luttinger effect.

To this end we consider the effective Hamiltonian, $H_{eff}$,
between holes on the hexagonal backbone to next-leading order in 
$t/J^z$,
induced by excursions to the surrounding sites in which the system 
leaves the ground state manifold of $H_J$. 
The matrix elements of this
effective Hamiltonian are determined by perturbation theory as
follows. First, we define a $k$-hole state with holes at sites
$m_1,m_2, ...\ m_k$\ (with some ordering of the sites, and for
concreteness with $m_i<m_j$\ for $i<j$) as 
\bea \left|
m_1 m_2\ ...\ m_n\right>= 
\prod_{i=1}^{k}c_{i\uparrow} \left| G\right>.  
\eea 
In the two-hole sector ($k=2$) we thus obtain: 
\bea \left<m_1
m_2\right| H_{eff}\left|m_1 m_2\right> = \sum_{p_1,p_2}\frac{\left|
\left< p_1 p_2 \left| H_t \right| m_1 m_2\right> \right|
^2}{E^0_{m_1,m_2}-E^0_{p_1,p_2}}; 
\eea here $E^0$\ are the energies of
the two-hole states for $t=0$. Note that the overall sign
of the many-hole states is immaterial at this order as only the
squares of matrix elements of $H_t$\ are used in determining 
$H_{eff}$.

To this order, more general, non-diagonal matrix elements are absent
since holes hopping of the backbone have to hop back to their original
site in order to reconstitute a leading-order ground state. 
When the two holes
are not on the same hexagon, $H_{eff}$\ always takes the 
same value which
is just twice the contribution of an isolated hole and hence has the
interpretation of an effective one-body potential as 
discussed in the last
section. Subtracting this leaves a term that depends on the joint
presence of the two holes and can thus be viewed 
as an effective two-body 
potential, $V(R)$, for the holes. This has the form, 
restoring $\eta$ for
future use:
\bea
V(R= 1)&=&\infty \nonumber\\
V(R= \sqrt{3})&=&-\frac{t^2}{J}
\frac{2(13-4\eta-\eta^2)}{3-4\eta+\eta^2} 
\nonumber\\
V(R= 2)&=& -\frac{t^2}{J}\frac{2(1+\eta)}{3-\eta} 
\nonumber\\
V(R> 2)&=&0 \nonumber,
\eea
where $R$\ is measured in lattice constants of the triangular 
lattice.

This effective interaction is dominantly repulsive on account of the
hard core piece considered earlier. Hence the additional, weak,
attraction can only induce a superconducting instability of the
parent Fermi liquid and we do not need to worry about the possiblity
of phase separation at $\eta=0$. In the continuum, it is easy to
see that the two body problem with a hard core and a weak attractive
tail posesses attractive phase shifts in sufficiently high angular
momentum channels---at low densities this is sufficient to lead
to pairing.\cite{emery} While we have not carried out a detailed 
analysis on the hexagonal lattice, the general conclusion will hold.

We end this section by considering the effect of switching on an
$\eta>0$. For $\eta<1/3$, the above arguments retain their validity,
with the main consequence of $\eta$\ being a strengthening of the
attractive part of the effective interaction. For $\eta>1/3$, the
allowed many-hole states change entirely in character. The
density-density attraction $H_\eta$\ then is strong enough to overcome
the exchange term $H_J$\ and produce phase separation. This happens
because phase-separation costs an energy of $J(3\eta+1)/4$\ per hole,
whereas the optimal states described above achieve a cost of
$3J\eta/2$.

\section{Effective interaction for classical holes}

Thus far we have focussed on the physics precisely at $T=0$. In this
limit the degeneracy of the parent magnet is ``selected away''. At
non-zero temperatures, but still below $J^z$, the ground-state entropy
of the parent magnet will again play a role. As a first step towards
an understanding of finite temperatures we consider $t \ll T \ll J$,
equivalently, we discuss the behaviour of classical annealed holes on
the triangular antiferromagnet.\cite{fn-quenchedholes}
 Our central result in this part is an
effective interaction, induced entropically, between two holes. We
close with some remarks on the finite density problem.

\subsection{Two holes}

Two holes 
on the triangular lattice experience an entropic 
interaction because the number of ground-state spin configurations, 
$Z({r})$, minimising $H_J$\ depends on their separation vector, 
${r}$. From a knowledge of $Z({ r})$, upon subtracting
the ``creation entropy'' of the two holes, $\ln Z(\infty)$, we can
obtain an effective interaction potential $\beta v(r)$, 
which vanishes as 
$r\rightarrow \infty$:
\bea
\beta v(r)&=&-\ln Z(r)+\ln Z(\infty)\nonumber\\
&=&-\ln (Z(r)/Z) +\ln (Z(\infty)/Z);
\eea
here $Z$\ is the partition function of 
the undiluted system.\cite{fn-dimerholes} 

Defining $Z(\infty)/Z\equiv\fl^2$\ and
$Z(r)/Z\equiv\fl ^2-\zeta(r)$, we obtain
\bea
\beta v(r)=-\ln\left[ 
1-\frac{\zeta(r)}{\fl^2}\right]\simeq\frac{\zeta(r)}{\fl^2},
\eea
the last step being justified at large $|r|$\ by  the smallness of
$\zeta(r)/\fl^2=0$, as we will show below.

To determine $\zeta(r)$\ and $\fl$, it is convenient to use a standard
representation of the spin problem as a classical dimer model on the
dual hexagonal lattice.  The presence of a hole is encoded by a
certain dimer configuration around it.  The latter is soluble by the
Pfaffian techniques introduced by Kasteleyn \cite{kasteleyn61}
although we will find it convenient to use the language of Grassmanian
functional integrals introduced into such problems by
Samuel.\cite{samuel} Thus, the function $\zeta(r)$\ is obtained by
evaluating a twelve-fermion correlation function.

In detail, this calculation proceeds as follows.
First, we map ground state spin configurations on the triangular lattice to 
dimer configurations on its dual hexagonal lattice by
drawing a dimer through each frustrated bond, placing its endpoints at
the centres of the two triangles sharing the frustrated bond. Noting
that each triangle has one and only one frustrated bond in the ground
state, the dimers thus provide a hard-core covering of the lattice
dual to the triangular lattice, namely the hexagonal
lattice. Conversely, up to the global Ising symmetry, each dimer
covering corresponds to a ground state of the Ising model.

The statistical mechanics of classical hard core dimer coverings on
the hexagonal lattice has been studied by Kastelyn\cite{kasteleyn63}
and some correlation functions have been computed by Yokoi, Nagle and
Salinas.\cite{yokoi86}

\begin{figure}
\epsfxsize=1.5in
\centerline{\epsffile{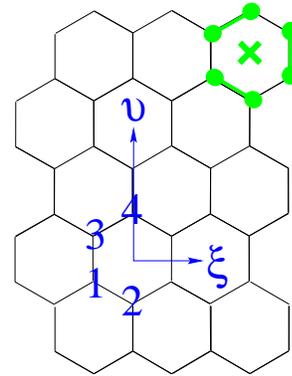}}
\caption{ 
The hexagonal lattice with the spins of the unit cell labelled 1-4. 
The lattice translation vectors are denoted by $\xi$\ and 
$\upsilon$\ 
for the $x$\ and $y$\ directions, respectively. 
The plaquette on the top right encloses a hole, marked by a cross. 
One of
the two possible 
resulting dimer plaquette configurations is shown.
}
\label{fig:pfaff}
\end{figure}

We extend this method to include the presence of diluted sites on the
original triangular lattice.  As outlined above, a hole can only
occupy the site of a spin experiencing a net exchange field of
zero. In dimer language, this means that only a spin at the centre of
a hexagonal plaquette occupied by exactly three dimers can be replaced
by a hole, as depicted in Fig.~\ref{fig:pfaff}. To determine $Z(r)$,
we are thus only allowed to count those dimer configurations in which
such plaquettes are encountered at the location of the holes.

To calculate $Z$\ in the first instance, one defines Grassman
variables, $\psi_i$, on each site of the $s$\ sites on the hexagonal
lattice. One then writes 
\bea Z=\int \left(
\prod_{l=s}^{1}d\psi_l\right) \exp(\psi_i A_{ij}\psi_j)={\rm Pf}(A)
=\sqrt{\det A},
\eea
where Pf$(A)$\ is the Pfaffian of the (antisymmetric) matrix $A$. 

The problem of determining $A$\ for a general class of two-dimensional
lattices has been solved by Kasteleyn. For the hexagonal lattice,
which we are interested in, it turns out that one has to double the
unit cell of the lattice to contain 4 spins, with the unit cell
labelled as in Fig.~\ref{fig:pfaff}, so that $A_{ij}(r)$\ equals
\bea
\left( 
\matrix{
0&\de{r,0}-\de{r,\xi}&\de{r,0}&0\cr
-\de{r,0}+\de{r,-\xi}&0&0&\de{r,\upsilon}\cr 
-\de{r,0}&0&0&\de{r,0}-\de{r,\xi}\cr
0&-\de{r,-\upsilon}&-\de{r,0}+\de{r,-\xi}&0\cr}
\right).\nonumber
\eea
Here, the vectors $\xi$\ and $\upsilon$\ denote the 
translation vectors of the lattice with the doubled unit cell 
(Fig.~\ref{fig:pfaff}).
The holes
are located in cells $r_1$, site $i$, and $r_2$, site $j$, 
with $r=r_2-r_1$.

This choice of $A$\ accomplishes the following. If the exponential in
the integrand for the partition function is expanded, the only terms
that result in a nonzero contribution to the integral are those in
which each $\psi_i$\ appears exactly once. Each such term represents a
covering of dimers with sites $i$\ and $j$\ being connected by a dimer
if $A_{ij}$\ is one of the prefactors of the term. The signs of the
entries of $A$\ are chosen such that every such term in fact
integrates to 1 so that all dimer coverings are allocated equal
weight.  One thus obtains ${\cal S}/k_B=\ln Z/N\simeq 0.323$.

We first caclulate the probability that a spin can be replaced with a
hole, which we require as this will turn out to be $\fl$, defined
above. To ensure the plaquette surrounding a spin (the sites of which
we label $1 ... 6$) is covered by three dimers, we insert a prefactor
$A_{12}A_{34}A_{56}\psi_1\psi_2\psi_3\psi_4\psi_5\psi_6$\ into the
integrand, which forces the three dimers into the required
position. 
We thus need to calculate
\bea
\fl&=&\frac{1}{Z}\int \left(
\prod_{l=s}^{1}d\psi_l\right) \times \nonumber\\
&&\ \ \ A_{12}A_{34}A_{56}
\psi_1\psi_2\psi_3\psi_4\psi_5\psi_6\exp(\psi_i A_{ij}\psi_j).
\eea

Since the Fermionic action is quadratic, this becomes by Wick's
theorem the sum over all $6!$\ contractions of pairs of $\psi$s: 
\bea
\fl&=&- \sum_{P} {\rm sign}(P) G_{i_1i_2}G_{i_3i_4}G_{i_5i_6}\ , 
\eea
where the sum runs over all permutations 
$P=\{ i_l|l=1 ... 6\}$\ of $\{
1 ... 6\}$. These Green functions have been worked out by Yokoi
\etal\cite{yokoi86} together with identities 
expressing all $G_{ij}$\
in terms of $G_{24}$, which for $y\geq0$\ is given by
\bea
G_{24}(x,y)
&=&
\frac{2}{\pi}(-1)^x\int_0^{\pi/3}d\phi 
\frac{\cos(2x\phi)}{\left|1+\exp(2i\phi) \right|^{2(y+1)}},
\eea
with $G_{24}$\ given by the negative of this expression for $y<0$.

In fact, the sum for $\Upsilon$\ only contains $3!$\ nonzero terms as
Green functions arising from contractions of sites belonging to the
same sublattice of the hexagonal lattice vanish. We thus obtain 
\bea
\fl= \frac{2}{27} - \frac{3\, {\sqrt{3}}}{8\, {\pi }^3} +
\frac{1}{{2\sqrt{3}}\, \ \pi } \simeq 0.14501.  
\eea 
To confirm the
calculation so far, we remark that Monte Carlo simulations we have
carried out in a different context (frustrated transverse field Ising
models)\cite{mcs2000} give $\fl=0.147\pm0.003$.

The evaluation of $Z(r)$\ thus requires twelve-fermion correlation
functions as two holes have to be introduced. There are two technical
difficulties in this calculation. Firstly, the distance-dependence of
the Green functions is not known in closed form, and secondly, we now
require $6!=720$\ terms each containing at least six Green functions
even after using the vanishing of same-sublattice Green functions.

\begin{figure}
\epsfxsize=3in
\centerline{\epsffile{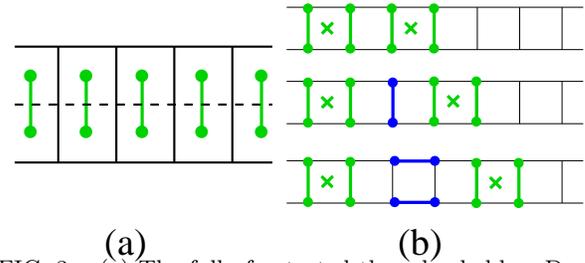}}
\caption{ 
(a) The fully frustrated three-leg ladder. 
Dashed (solid) lines indicate antiferromagnetic (ferromagnetic) 
exchange. The dimer state corresponding to all spins pointing up
is shown.
(b) The corresponding dimer model on the dual lattice. 
A hole, marked by a cross, fixes the dimer plaquette surrounding it.
Odd hole separations are entropically unfavourable because dimers 
can only resonate in pairs (bottom graph) and not on their own
(middle graph).
}
\label{fig:tlld}
\end{figure}

As the reader can see, this calculation is straightforward in
principle but somewhat involved in practice---it will turn out though
that the final answer has a transparent rationale. Before proceeding
further it is useful to discuss the same general phenonmenon in a toy
model which is exactly solvable and where the combinatorics is more
transparent.  Consider a one-dimensional dimer model in which the
dimers cover a two-leg ladder (see Fig~\ref{fig:tlld}).  The entropy
per rung of this dimer model is the golden mean,
$G=(\sqrt{5}+1)/2\simeq 1.62$.  The dimer model on this ladder arises
in the study of a fully frustrated three-leg Ising
ladder,\cite{rmslsprep} and removing a spin from this Ising model
corresponds to forcing the square surrounding this spin to be covered
by two parallel dimers.

If two holes are adjacent to one another (Fig.~\ref{fig:tlld}b, top),
there are no rungs separating the plaquettes surrounding them. Moving
the hole on the right two units further away (Fig.~\ref{fig:tlld}b,
bottom) costs an entropy of $\ln G^2\simeq \ln 2.62$\ in the
semi-infinite region outside to the right of the hole pair and only
gains an entropy of $\ln 2$\ as the dimer pair between the holes can
only have two states. For even separations, it is thus advantageous
for the holes to move close together.

Separating the holes by one unit costs entropy $\ln G$\ outside while
gaining nothing inside as a single dimer can only exist in the
vertical state (Fig.~\ref{fig:tlld}b, middle). 
Moving the hole further
away by two units, however, yields a total gain of $\ln 3-\ln G^2>0$\
in entropy as the three dimers between the hole plaquettes can have
three states. Therefore, for odd separations, the holes want to move
further apart. In fact, the alternating two-hole potential for this
model can be calculated both by Pfaffian and transfer matrix methods,
and takes the form
\bea 
\beta v_{\rm ladder}(r)\sim
\frac{\zeta_{\rm
ladder}(r)} {\Upsilon^2_{{\rm ladder}}}
=\frac{4}{5}\frac{G-1}{G^3\Upsilon^2_{{\rm ladder}}}
(-1)^{r+1}G^{-2r},
\eea with $\Upsilon_{{\rm ladder}}=1/\left(G\left(2G-1\right)
\right)$.

Returning to the triangular lattice problem, the large $|r|$\
asymptotics
of the Green functions are calculated in Ref.~\onlinecite{yokoi86}:
\bea
G_{24}(x,y)\simeq{\left( -1 \right) }^x\, \frac{
     {\sqrt{3}\,y\,\cos (\frac{2\,\pi \,x}{3})}
       {} + {x\,
         \sin (\frac{2\,\pi \,x}{3})}  }{
    {{\pi }\,x^2} + 
      3\, {\pi }\,y^2 } \ .
\eea

As these Green
functions decay algebraically with $|r|$, the terms can be arranged
according to how many contractions between sites belonging to
different holes they contain. The leading terms are the ones in which
all contractions are between sites of one or the other hole, and these
terms are clearly independent of $r$. In fact, it is easy to see that
the contractions around a given hole are simply those which yielded
$\fl$\ in the single-hole calculation, and thus the leading,
$r$-independent term is $Z(\infty)/Z=\fl^2$, as promised above.

The leading non-trivial $r$-dependence arises from terms which contain
two Green functions connecting sites around the two holes, an odd
number of connections being impossible. There are 324 such terms. With
the assistance of Mathematica, we were able to analyze their
asymptotics and obtain the exact long distance form,
\bea 
\zeta(x,y)\sim -\Xi \frac{\cos(4\pi x/3)}{x^2+3y^2}\ ,
\eea 
where 

\bea
\Xi&=&\frac{729 - 324\,{\sqrt{3}}\,\pi  - 324\,{\pi }^2 + 
    96\,{\sqrt{3}}\,{\pi }^3 + 64\,{\pi }^4}{288\,{\pi }^6}
\nonumber\\
&\simeq& 0.025853.
\eea
Translating the co-ordinates back to the original triangular
lattice, our final asymptotic 
expression for the interaction potential
is
\bea
\beta v(r) \sim -\frac{\Xi}{\fl^2} \frac{\cos(4\pi x/3)}{x^2+y^2}\ .
\label{eq:finalint}
\eea

The physics of this expression is quite transparent. The wavevector
${\bf q} = {4 \pi \over 3} \hat{x}$\ on the triangular lattice is
characteristic of the hexagonal frustrated bond pattern sketched in
Fig~\ref{fig:hex}b. It has long been recognized (already by Wannier
\cite{wanhout}) that this pattern represents the dominant ordering
tendency in the ground state average of the classical problem. In a
more modern formulation in terms of height representations
\cite{blote} the pattern is the reference flat surface about which the
system exhibits critical fluctuations.  In accordance with this we
find from Eq.~\ref{eq:finalint} that the holes attract when they share
a hexagonal backbone of frustrated bonds, and repel when they do
not. The algebraic decay could have been anticipated in the dimer
language from our observation that the leading piece at large distance
comes from a sum of dimer-dimer correlation functions each of which
individually decays as the inverse-square of the distance. However,
and this is what motivated the computation, it is not evident in
advance that the various dimer correlators (which enter with different
signs) do not sum to zero and hence the acutal result does not just
yield a non-trivial constant but also assures us that it is not zero.

\subsection{Finite density of classical holes: an incompressible 
state}

As the two hole interaction derived previously has considerable
spatial structure and a long range, the reader may wonder whether it
leads to a condensation of a finite density of holes on the hexagonal
backbone. This does not happen.  To see this note that $\beta v(r)$ is
{\it independent} of $T$, which is to say that there are flucutations
even as $T \rightarrow 0$---as indeed there are on account of the
degeneracy.  In the dilute limit, in which the two hole calculation is
reliable, the effective interaction strength is proportional to the
inverse square of the hole separation, i.e. to the (low) density and
hence the hole gas remains in a liquid phase. Nevertheless, the long
range of the interactions indicates that it exhibits substantial
correlations coming from the entropic potential, and---on account of
the form of the interaction---we suspect that such correlations will
be algebraic as well. As we have already argued that the zero
temperature state involves hole condensation on the hexagonal backbone
driven by the hole kinetic energy, it appears that this phenomenon
arises for both energetic and entropic reasons, and is thus quite
robust so that it will characterize the entire temperature range $T
\ll J^z$. (In all of this we are again referring to $\eta < 1/3$, else
the energetic attraction ignored in the entropic derivation will again
dominate and drive phase separation.)

Finally we report a curiosity regarding classical holes for
$\eta<1/3$.  As the density of holes is increased, it is not clear
whether the interaction finally becomes strong enough to produce phase
separation into regions either fully occupied or maximally diluted
(but respecting the hard core repulsion of the holes). However, even
if it does not, there exists a density of holes at which long-range
order is present.
 
The closest packing compatible with the ground-state constraint is
obtained by placing the holes such that they cover one of two
sublattices of the hexagaonal backbone of up spins depicted in
Fig.~\ref{fig:hex}b.  At a hole density of 1/3, this sublattice is
completely occupied by holes, at zero energy cost at $O(J)$\ compared
to the undiluted system.

Adding further holes above this density costs a magnetic energy of
order $J$, and thus the chemical potential experiences a discontinuity
at a hole density of 1/3.  The state at this density, which has
perfect unfrustrated antiferromagnetic order on the bipartite
hexagonal lattice occupied by the remaining spins, is thus an {\em
incompressible} state.

As the density of holes is increased further, phase separation 
occurs for certain.  Completely depleted regions are created by
removing one hexagonal ring after another as the system optimises its
energy by removing those spins with the smallest number of
neighbours. This is the same mechanism that would operate in any
unfrustrated magnet. As mentioned above, for $\eta>1/3$, phase
separation occurs for all hole densities.

\section{Summary}

We have considered the problem of dilute, slow holes in the triangular
Ising antiferromagnet. We find that doping has two very different
outcomes depending on the dimensionless strength $\eta$ of the
density-density interaction. For $\eta < 1/3$ we find quite generally
that at all temperatures $T \ll J^z$ the holes tend to condense on a
hexagonal backbone of frustrated bonds and that at $T=0$ they form a
superconducting state that coexists with magnetic order and the
breaking of translational symmetry in a ``magnetic supersolid''.
As a special case of this more general phenomenon, we were able to 
prove a ``Nagaoka'' theorem for a single hole at $J^z = \infty$. 
For $\eta > 1/3$ in our short ranged model, we find phase separation. 
Whether the inclusion of
the long range piece of the Coulomb interaction will change that
outcome and stablize the magnetic supersolid and how much of this
structure will persist to larger values of $t/J^z$ remain topics for
future work.

\section*{Acknowledgements}
We are grateful to D. Priour and P. Chandra, who also kindly commented 
on the manuscript, for collaboration on
closely related work. We would also like to acknowledge A. Yodh
for a stimulating talk on entropic forces that led us to examine
the same question in the context of this paper.
This work was supported in part by grants from the Deutsche
Forschungsgemeinschaft, NSF grant No. DMR-9978074, the A. P. Sloan
Foundation and the David and Lucille Packard Foundation.



\begin{references}


\bibitem{pwa87}
P. W. Anderson, \tit{Science}{235}{1196}{1987}{THE RESONATING VALENCE
BOND STATE IN LA2CUO4 AND SUPERCONDUCTIVITY}.

\bibitem{br} W. F. Brinkman and T. M. Rice, Phys. Rev. B {\bf 39},
6880 (1970); L.~N.~Bulaevskii and D.~I.~Khomskii,
Zh. Eksp. Teor. Fiz. {\bf 52}, 1603 (1967) [Sov. Phys. JETP {\bf 25},
1067 (1967)]; Fiz. Tverd. Tela {\bf 9}, 3070 (1967) [Sov. Phys. Solid
State {\bf 9}, 2422 (1968)]
 

\bibitem{fn-weak} This argument is compelling at weak doping. At
sufficiently high dopings one might anticipate a renormalized
fermi liquid description, e.g. as assumed in the work on the
spin fluctuation scenario for the cuprates. See e.g. D. Pines, 
cond-mat/0002281 and references therein.

\bibitem{rvb} G. Baskaran, Z. Zou and P. W. Anderson, Solid
State Commun. {\bf 63}, 973 (1987); D. S. Rokhsar and S. A.
Kivelson, Phys. Rev. Lett. {\bf 61}, 2376 (1988); 
P. A. Lee, Physica C 317-318, 194 (1999) and references therein.


\bibitem{stripes}
Three entirely different account of stripe formation can be found
in J. Zaanen and O. Gunnarsson, Phys. Rev. B {\bf 40}, 7391 (1989);
U. L\"ow, V. J. Emery, K. Fabricius and S. A. Kivelson, Phys.
Rev. Lett. {\bf 72}, 1918 (1994) and M. Vojta and S. Sachdev,
Phys. Rev. Lett. {\bf 83}, 3916 (1999).

\bibitem{ps} V. J. Emery, S. A. Kivelson and H. Q. Lin,
Phys. Rev. Lett. {\bf 64}, 475 (1990). 

\bibitem{nagaoka} Y. Nagaoka, Phys. Rev. {\bf 147}, 392 (1966).

\bibitem{wanhout} 
G. H. Wannier, \jour{\pr}{79}{357}{1950};
R. M. F. Houtappel, \jour{Physica}{16}{425}{1950}.

\bibitem{mcs2000} 
R. Moessner, S. L. Sondhi and P. Chandra,
\tit{\prl}{84}{4457}{2000}{Two-dimensional periodic frustrated Ising
models in a transverse field}.

\bibitem{fn-tgoestozero} Perhaps the best way to set up the contrast
is to consider the limit $T\rightarrow 0$ for the insulator and with a
single hole present. In the former case we obtain the symmetric ground
state manifold average, while in the latter we get and average over
the state $\left. |h \right>$ and its symmetry related counterparts.

\bibitem{fn-contrast} Readers familiar with the Nagaoka theorem may
worry about this point. In the Hubbard problem, reducing $U$ from
infinity, where the theorem holds, has a dramatic effect in that it
confines the polarization to a finite region surrounding the hole.
There, this happens since a finite $U$ permits virtual processes far
from the hole which generate an antiferromagnetic exchange. The
resulting exchange strength is $O(t^2/U)$\ but it exacts a volume cost
should we attempt to keep the system polarised. In our problem, single
occupancy is present at the first step and, far from the hole, {\it
no} virtual processes are possible, whence no volume cost arises.

\bibitem{fl}
For the more general case of fermions with spin, the two dimensional
low density problem has been studied by P. Bloom, Phys. Rev. B {\bf 12},
125 (1975) and by J. R. Engelbrecht and M. Randeria, Phys. Rev. Lett.
{\bf 65}, 1032 (1990).

\bibitem{kohn-lutt} W. Kohn and J. M. Luttinger, Phys. Rev. Lett.
{\bf 15}, 524 (1965). For spinless fermions see the renormalization
group analysis of R. Shankar, Physica A {\bf 177}, 530 (1991).

\bibitem{emery}
V. J. Emery, \tit{\np}{12}{69}{1959}{On the existence of solutions
of the Brueckner equations for a many-fermion system}; 
\tit{\np}{19}{154}{1960}{Reaction matrix singularities and the 
energy gap in an infinite system of fermions}. 

\bibitem{fn-quenchedholes} Complementary to the study of {\it
annealed} dilution, there exists---under the general heading of `order
by disorder'---a body of work on the role of {\it quenched} dilution
in lifting the ground-state degeneracy of frustrated magnets,
beginning with J. Villain, R. Bidaux, J. P. Carton and R. J. Conte,
\tit{J. Physique}{41}{1263}{1980}{Order as an effect of disorder}.

\bibitem{fn-dimerholes} We note that this problem is quite different
from that of introducing holes (monomers) on the dimer lattice itself.
That latter leads to a confining potential that grows with distance,
as shown by M.E. Fisher and J. Stephenson, Phys. Rev. {\bf 132}, 1411
(1963).

\bibitem{kasteleyn61}
P. W. Kasteleyn, 
\tit{Physica}{27}{1209}{1961}{The statistics of dimers on a lattice}.

\bibitem{samuel}
S. Samuel, \tit{J. Math.\ Phys.\ }{21}{2806}{1980}{The use of 
anticommuting variable integrals in statistical mechanics. I. 
The computation of partition functions.}



\bibitem{kasteleyn63}
P. W. Kasteleyn, \tit{\jmp}{4}{287}{1963}{Dimer statistics and 
phase transitions}

\bibitem{yokoi86}
C. S. O. Yokoi, J. F. Nagle and S. R. Salinas, 
\tit{J. Stat.\ Phys.\ }{44}{729}{1986}{Dimer pair correlations 
on the Brick lattice}.

\bibitem{rmslsprep}
R. Moessner and S. L. Sondhi, {\em in preparation}.

\bibitem{blote} 
B. Nienhuis, H.J. Hilhorst and H.W.J. Blote,
\tit{J.\ Phys.\ A}{17}{3559}{1984}
{TRIANGULAR SOS MODELS AND CUBIC-CRYSTAL SHAPES}.

\end{references}
\end{document}